\documentclass[useAMS,usenatbib]{mn2e}
\usepackage{graphicx}
\usepackage{graphics}
\usepackage{rotating}
\usepackage{amsmath}
\usepackage{amssymb}

\usepackage{times}

\newcommand{\sng}{G1.9+0.3}
\newcommand{\mjyb}{mJy beam$^{-1}$}
\newcommand{\changes}[1]{{#1}}

\title[A 20 Year Radio Light Curve for \sng]
      {A 20 Year Radio Light Curve for the Young Supernova Remnant G1.9+0.3}
\author[T. Murphy et al.]
       {T. Murphy$^{1,2}$\thanks{E-mail: tara@physics.usyd.edu.au},
        B. M. Gaensler$^{1}$,
        S. Chatterjee$^{1}$ \\
$^{1}$Institute of Astronomy, School of Physics, University of Sydney, NSW 2006, Australia\\
$^{2}$School of Information Technologies, University of Sydney, NSW 2006, Australia\\
}

\date{Accepted 0000 June 08. Received 0000 June 08; in original form 0000 June 08}

\pagerange{\pageref{firstpage}--\pageref{lastpage}} 
\pubyear{2008}

\begin{document}

\maketitle

\label{firstpage}

\begin{abstract}
The radio source \sng\ has recently been identified as the youngest
known Galactic supernova remnant, with a putative age of $\sim$100
years.  We present a radio light curve for \sng\ based on 25 epochs of
observation with the Molonglo Observatory Synthesis Telescope,
spanning 20 years from 1988 to 2007.  These observations are all at the same frequency
(843 MHz) and comparable resolutions ($43\arcsec \times 91\arcsec$ or
$43\arcsec \times 95\arcsec$) and cover one fifth of the estimated
lifetime of the supernova remnant.  We find that the flux density has
increased at a rate of $1.22 \pm {0.24 \atop 0.16}$ per cent yr$^{-1}$ over the last two
decades, suggesting that \sng\ is undergoing a period of magnetic field
amplification.
\end{abstract}


\section{Introduction}\label{s_intro}
The well known deficit of young supernova remnants (SNRs) in our Galaxy has motivated 
many searches for these objects \citep[for example;][]{green84,gray94d,misanovic02}.
Predictions based on extragalactic supernova (SN) rates suggest that there should be 
around 40 SNRs younger than 2000 years in our Galaxy \citep{cappellaro03}, and yet less than
10 have been identified \citep{green04}. The recent identification of \sng\ as a young 
SNR with age $\sim100$ years  by \cite{reynolds08} and \cite{green08}, adds to this set, 
and makes it potentially the youngest known Galactic supernova remnant.

Most Galactic SNRs detected at radio wavelengths are detected well after the initial 
supernova event ($> 10^3 - 10^4$ years). Cassiopeia A \citep[SN~1681$\pm$19;][]{fesen06} 
is the youngest SNR of known age, as $\sim330$ years. The early stages of radio supernova (RSN) 
evolution have been studied in a number of bright extragalactic sources; detailed studies
include,
SN~1979C in M100 \citep{weiler86}, SN~1980K in NGC~6946 \citep{weiler86,weiler92},
SN~1993J in NGC~3031 \citep{weiler07} and SN~1987A in the Large Magellanic Cloud 
\citep{ball01,manchester02}. However, the oldest RSN of known age 
is SN~1923A \citep{eck98}, which leaves a critical gap in our knowledge of supernova 
evolution at intermediate ages of $\sim50-300$ years. Hence the age estimate of 
$\sim100$ years for \sng\ makes it a useful probe of this period in SNR evolution.

\sng\ was first identified as a potential young SNR based on its small angular 
size of 1.2\arcmin\ \citep{green84}. Recent Chandra observations by \cite{reynolds08}
showed that the remnant had
expanded by $16\pm3$ per cent between earlier VLA observations in 1985 and the Chandra 
observations in 2007. This was confirmed by VLA follow-up observations by \cite{green08}
which gave an expansion of $15$ per cent over 23 years (or an expansion rate of $0.65$ per cent yr$^{-1}$).
This led to an age estimate of $\sim100$ years (or at most $150$ years) for \sng. 
\citet{green08} also use archival data from a variety of instruments (over frequencies
332 to 5000 MHz) to suggest that \sng\ has brightened by a rate of $\sim 2$ per cent per 
year over the same period.

We present 20 years of radio observations from the Molonglo Observatory Synthesis Telescope 
\cite[MOST;][]{mills81,robertson91}. These observations were carried out at constant
frequency (843 MHz) and comparable resolutions ($43\arcsec\times91\arcsec$ or
$43\arcsec\times95\arcsec$). They show that \sng\ has been increasing in brightness 
by a rate of $1.22\pm {0.24 \atop 0.16}$ per cent yr$^{-1}$ over this period. 
\changes{Our new estimate has the advantage that the
measurements were taken with the same instrument, whereas the \citet{green08} estimate was based on
observations from a range of instruments, compiled from the literature.}

\section{Observations and Data}
The observations presented have been carried out over the last 20 years 
with the Molonglo Observatory Synthesis Telescope.
The MOST operates 
at a central frequency of 843 MHz and has a restoring beam of 
$43\arcsec\times43\arcsec\csc|\delta|$ where $\delta$ is the central declination
of the field. Further technical specifications are in \cite{bock99}.

\sng\ is detected in a total of 30 archival fields covering the period 
1988 to 2007. These fields were observed as part of ongoing projects; 
the Molonglo Galactc Centre Survey \cite[1985--1991;][]{gray94a},
the Molonglo Galactic Plane Survey \cite[1983--1994;][]{green99a}, the second 
epoch Molonglo Galactic Plane Survey \cite[1996--2007;][]{murphy07}, and a 
Galactic centre monitoring program (2004--2007).
Discarding five fields in which the data quality was poor, we have 25
observations of \sng.
The field of view for the early observations was $70'$, which increased to
$163'$ after the MOST was upgraded in 1995. 
The restoring beam is  $43\arcsec\times95\arcsec$ for the pre-1995 fields and
$43\arcsec\times91\arcsec$ for the post-1995 fields.
Therefore the frequency and resolution were comparable for all of our 
observations, making this a self-consistent dataset covering approximately 
one fifth of the life of \sng.

The MOST data are reduced using a custom process described by \cite{bock99}.
The phase and flux density calibration are done by observing a set of calibrator
sources before and after each 12 hour observation. These calibrators
are strong, unresolved sources taken from the list given in \cite{campbell-wilson94}.
Analysis of the calibrator sources by \citet{gaensler00} showed the relative flux scale
(as measured by the average scatter in repeat observations of the non-variable calibrators)
to be accurate to around $3$ per cent. 
A similar analysis done by \citet{murphy07}, found the scatter for non-calibrator
sources observed by MOST to be around $5$ per cent.

The noise in individual MOST images is a combination of thermal noise and 
source confusion from both the main beam and sidelobes of the MOST. The noise
level varies with Declination. Since the telescope is calibrated only at the
beginning and end of each 12 hour run, there may be changes in the precise
operational condition of the telescope electronics and feeds that have a small
effect on the sensitivity of a given observation. The rms noise at the centre of
a typical MOST image is $\sim 1$ \mjyb. However, most of the fields containing
\sng\ have strong artefacts caused by the Galactic centre which is also in 
the field. Hence typical rms noise levels near the source are $\sim 50$ \mjyb. 

\section{Analysis}
\label{s_analysis}

\begin{figure*}
\begin{center}
\includegraphics[height=7.1cm,angle=270]{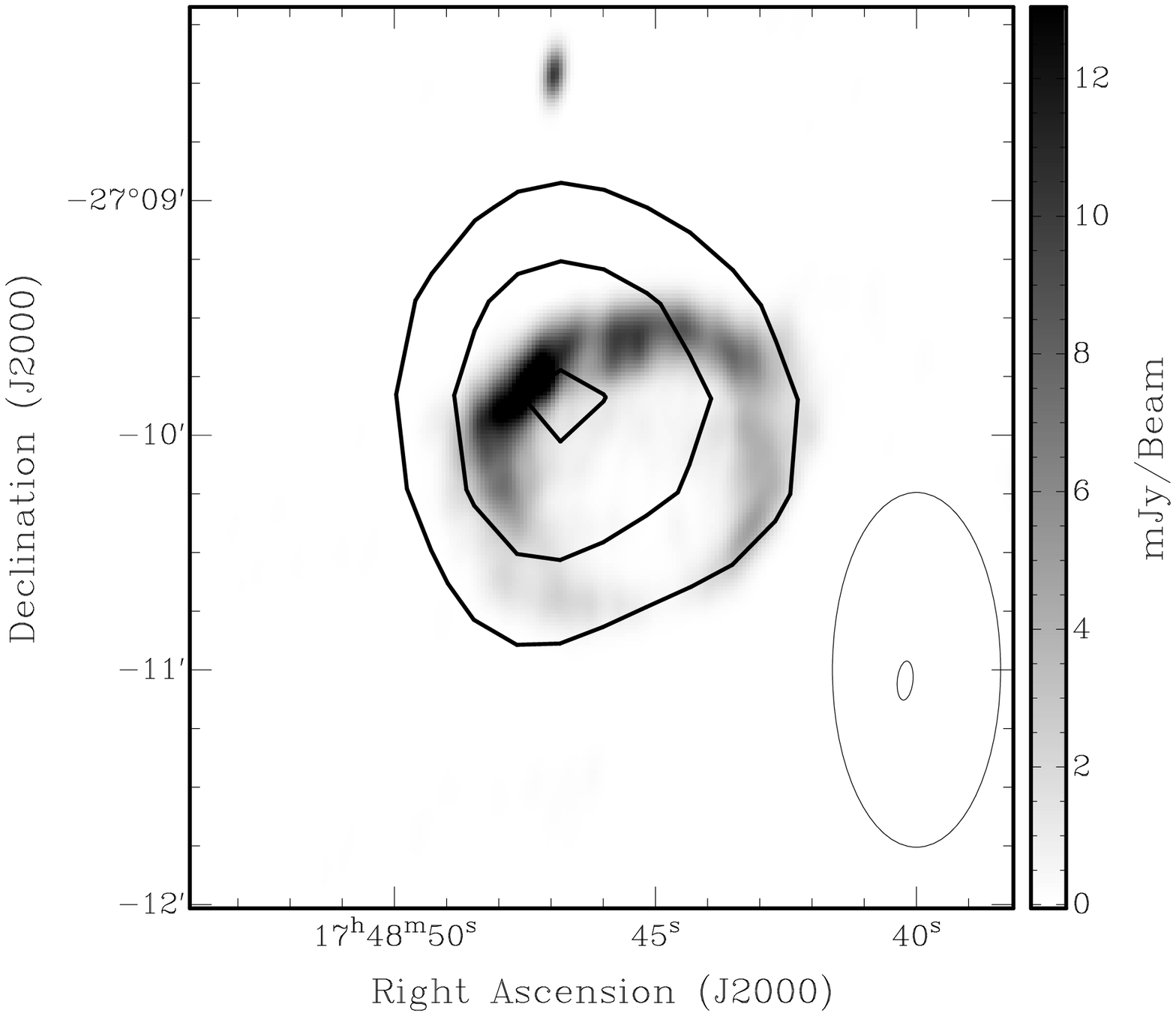}
\includegraphics[height=7.1cm,angle=270]{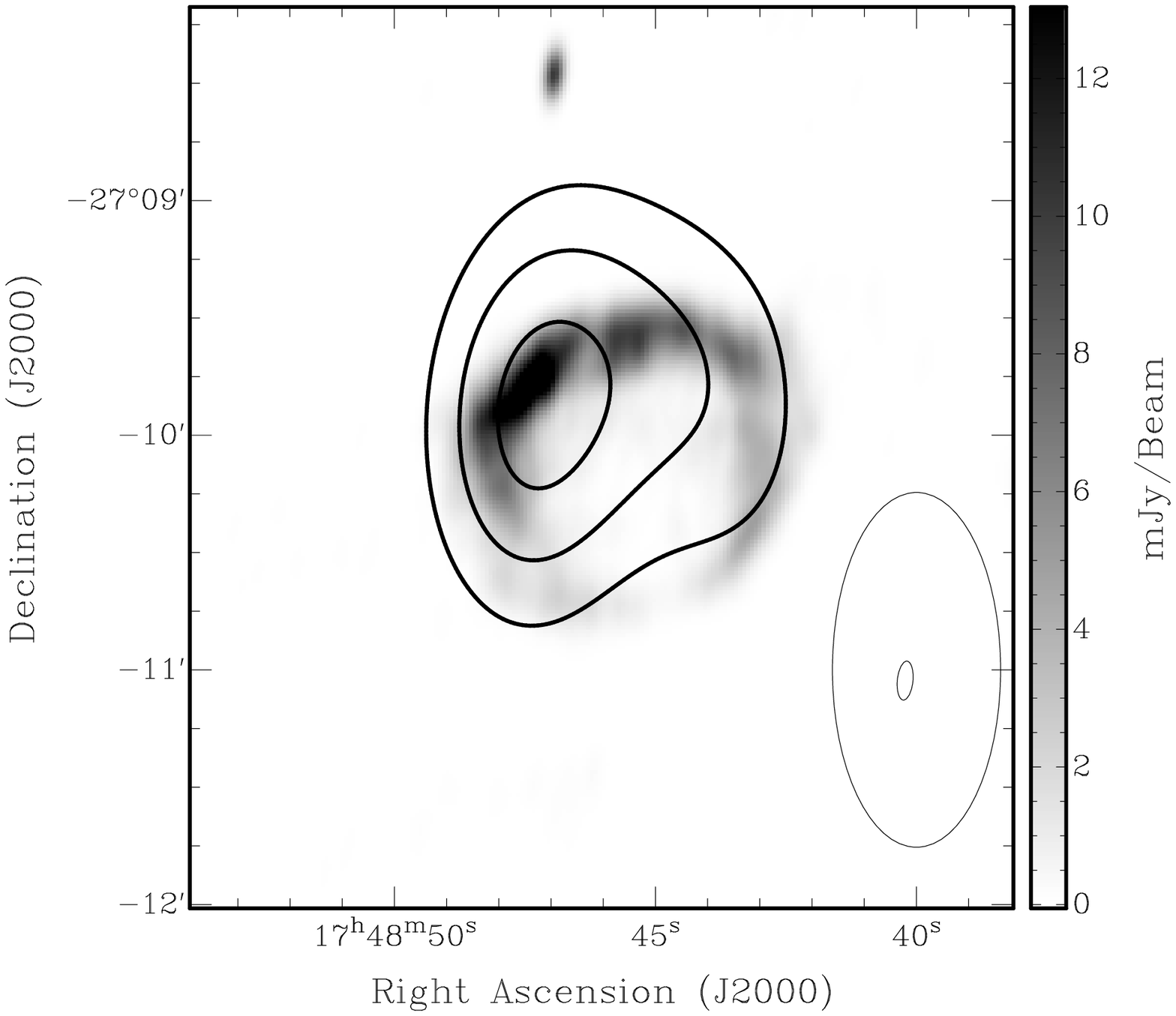}
\includegraphics[height=7.1cm,angle=270]{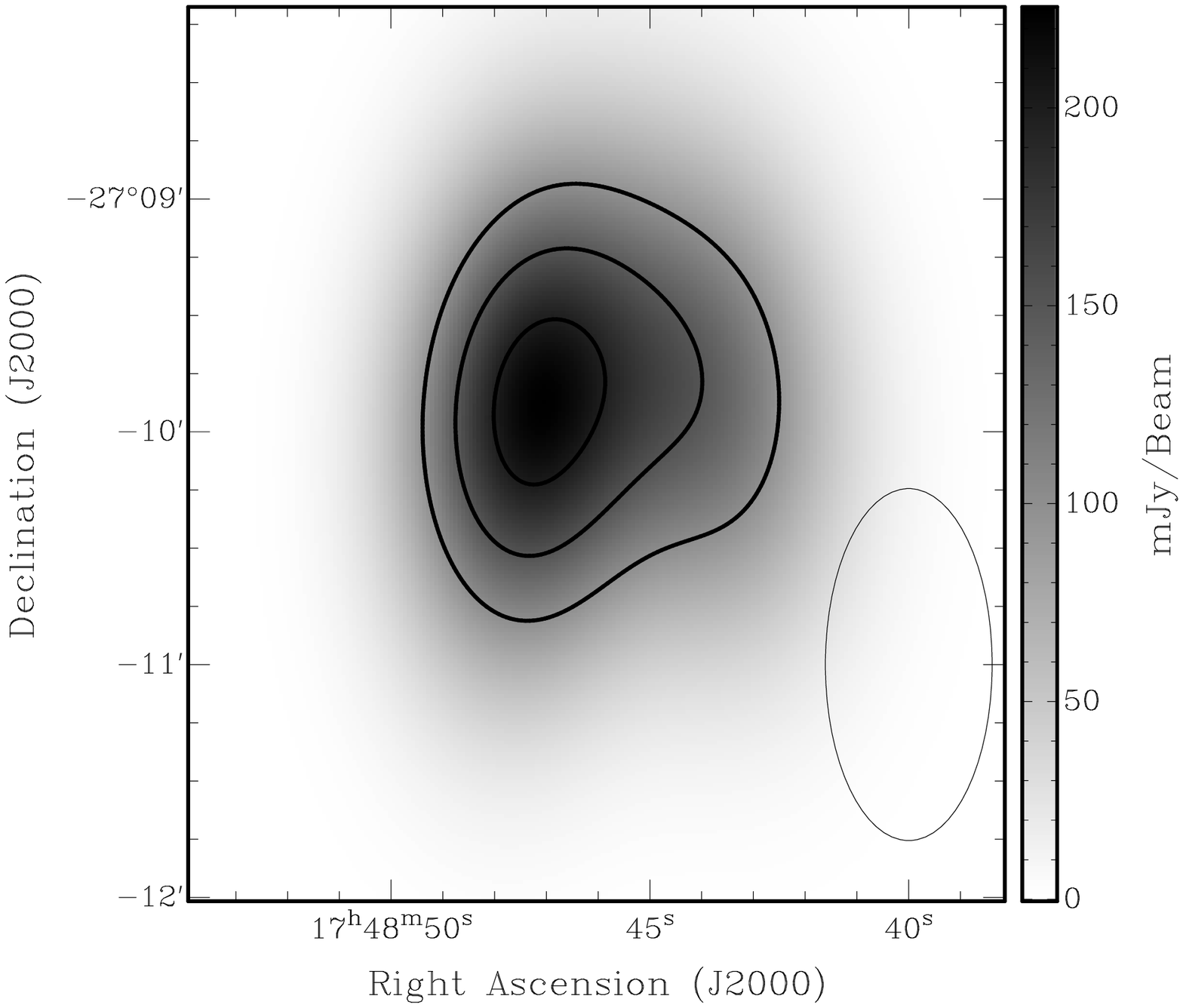}
\includegraphics[height=7.1cm,angle=270]{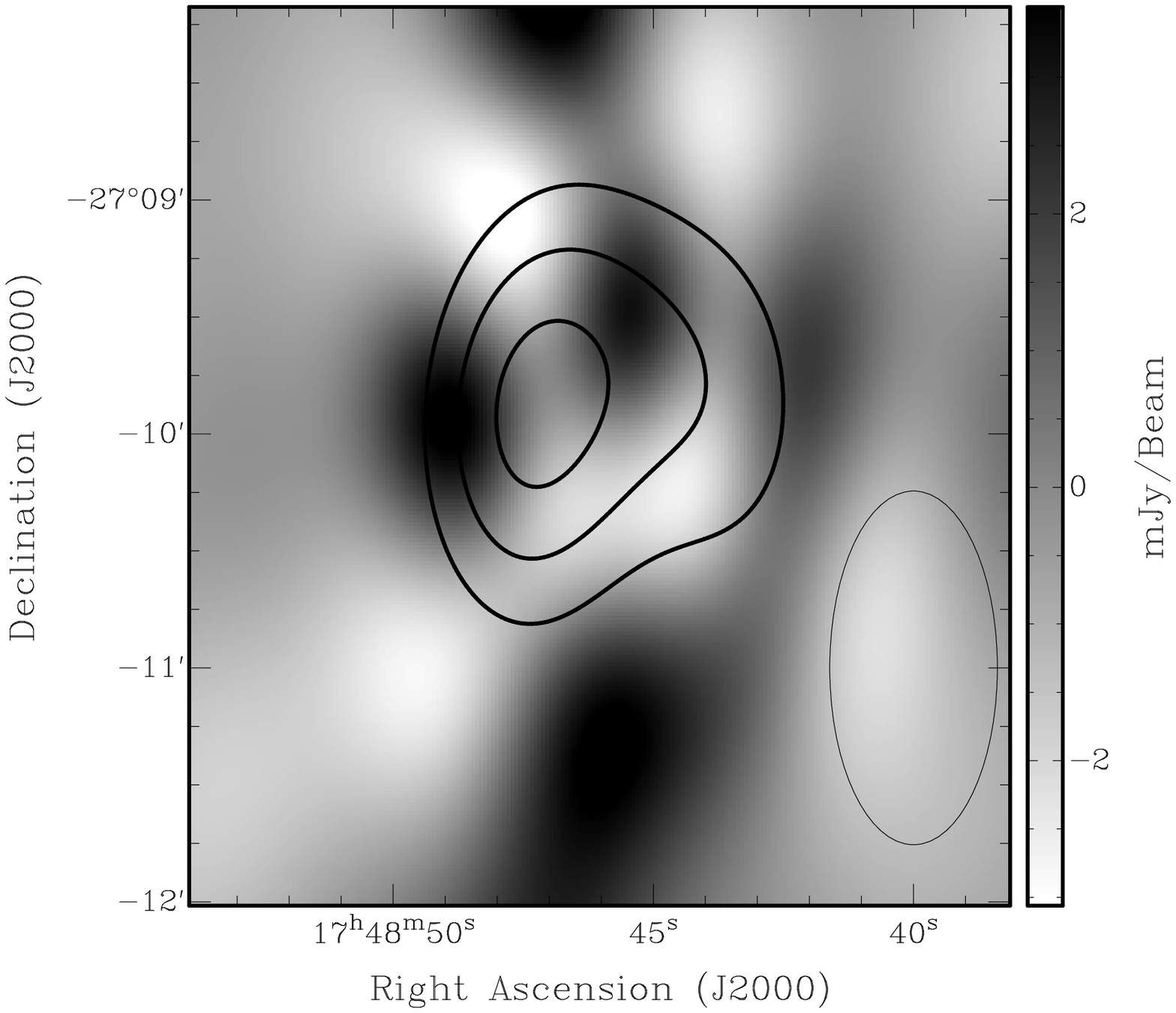}
\caption{A comparison of one MOST epoch (2007/06/19) with a VLA 4.8 GHz 
 image from \citet{green08}. {\it Top left:} MOST contours (200, 400 and 600 mJy 
 beam$^{-1}$) overlaid on VLA greyscale (range 0 to 13 \mjyb). The MOST and VLA beams
 are shown to the lower right. {\it Top right:} The VLA image smoothed to MOST 
 resolution is shown in contours (100, 150 and 200 \protect\mjyb) overlaid on the original image.
 {\it Bottom left:} A double Gaussian model fit is shown in greyscale (range 0 
 to 230 \protect\mjyb), with the smoothed VLA contours overlaid.
 {\it Bottom right:} The greyscale shows the residuals after
 subtracting the double Gaussian model fit from the smoothed VLA image 
 (range $-3$ to 3 \protect\mjyb) with the smoothed VLA contours for comparison.}\label{f_overlay}
\end{center}
\end{figure*}

\changes{
The MOST observations are not at high enough resolution to detect significant
changes in the size or mophology of \sng\ with time. \sng\ is only  
slightly extended ($\sim2$ beams E--W) in the MOST images, as shown in 
Fig.~\ref{f_postage} which shows a selection of our observed epochs.
}
Fig.~\ref{f_overlay} 
shows a MOST epoch (from 2007/06/19) overlaid on the 
\cite{green08} VLA 4.8 GHz image (from 2008/03/12). When the VLA image
is smoothed to the MOST resolution, there is good agreement in the observed 
morphologies.
The rms noise in the 4.8 GHz image is $\sim0.012$ \mjyb, and in the MOST image is 
$\sim30$ \mjyb.
\begin{figure*}
\begin{tabular}{cccc}
1988/07/06 & 1990/06/28 & 1997/06/26 & 1998/06/24 \\
\includegraphics[angle=270,width=3.2cm]{images/D1747263C.ps} &
\includegraphics[angle=270,width=3.2cm]{images/D1747262C.ps} &
\includegraphics[angle=270,width=3.2cm]{images/I1744283C.ps} & 
\includegraphics[angle=270,width=3.2cm]{images/I1744285C.ps} \\
1999/06/08 & 2004/05/27 & 2004/08/22 & 2005/01/22 \\
\includegraphics[angle=270,width=3.2cm]{images/I1744286C.ps} & 
\includegraphics[angle=270,width=3.2cm]{images/I1744288C.ps} & 
\includegraphics[angle=270,width=3.2cm]{images/I17442811C.ps} & 
\includegraphics[angle=270,width=3.2cm]{images/I17442813C.ps} \\
2005/10/22 & 2006/01/29 & 2006/10/07 & 2007/06/07 \\
\includegraphics[angle=270,width=3.2cm]{images/I17442819C.ps} &
\includegraphics[angle=270,width=3.2cm]{images/I17442821C.ps} & 
\includegraphics[angle=270,width=3.2cm]{images/I17442824C.ps} &
\includegraphics[angle=270,width=3.2cm]{images/I17442825C.ps} \\
\end{tabular}
\caption{MOST images for a selection of our observations, before the 
polynomial background subtraction. Contours are at 
50, 100, 200, 500 mJy beam$^{-1}$ and the greyscale range is
$-200$ to $-350$ mJy beam$^{-1}$. The MOST beam FWHM is shown in
the lower right of each image.}\label{f_postage}
\end{figure*}

\begin{figure}
\begin{center}
\includegraphics[width=\linewidth]{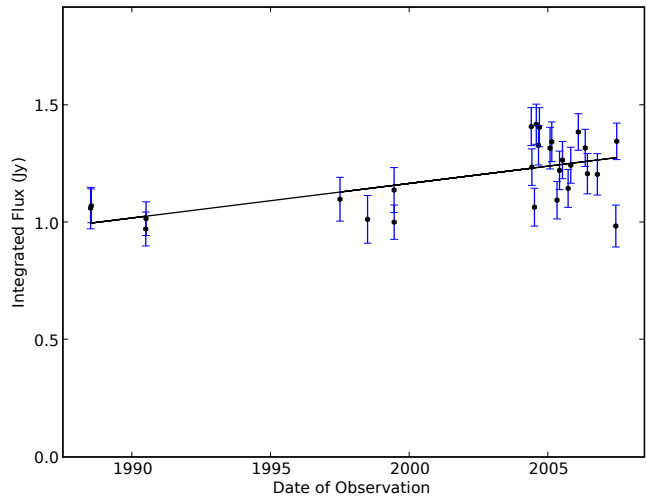}
\caption{843 MHz radio light curve for \protect\sng\ from 1988 to 2007. Flux densities were
calculated using a double Gaussian fit after polynomial subtraction of the background 
\changes{(see text for details of the flux density and error calculations)}. 
The solid line shows a 
least squares fit with gradient 0.015 ${\rm Jy}\;{\rm yr}^{-1}$ and a flux density of 
1.23 Jy on 2005/01/01.}
\label{f_curve}
\end{center}
\end{figure}

The integrated flux density of the source was measured by first fitting a 
two dimensional \changes{second order} polynomial surface to the surrounding region and 
subtracting it, 
and then fitting a double elliptical Gaussian to the source itself. Fitting the 
background was necessary due to the large flux gradient in the image caused by 
the artefacts from the Galactic center. In most of the fields the source is 
sitting in a significant negative trough that could have negative flux values 
as large as $-100$ mJy. 
Although the VLA observations show that morphology of the source is clearly a 
shell, at the MOST resolution it can be modelled reasonably well with a double elliptical 
Gaussian. This is demonstrated in the lower two panels of Fig.~\ref{f_overlay} which
show a double Gaussian model, and the residuals after subtracting the model from the 
VLA image smoothed to the MOST resolution. 
In the MOST images, the residuals after fitting were around the $\sim5$ per cent level, 
which we have included as a component of our overall error.
The resulting light curve is shown in Fig.~\ref{f_curve}. Errors on each point
are calculated by adding the estimated error in the flux density scale ($\sim5$ per cent), 
rms noise (calculated in region local to the source in each map, typically $20-50$ \mjyb\ or 
$\sim5$ per cent) and estimated fitting errors ($\sim5$ per cent) in quadrature.

Fitting a power law ($f\propto t^m$) to the light curve resulted in an exponent of $m=1.2$
with large errors.
A least squares fit of the light curve with a straight line gave a gradient of 
0.015~Jy~yr$^{-1}$, with a reduced $\chi^2$ of 2.0 for 25 degrees of freedom. 
Since this $\chi^2$ does not allow an accurate estimate of our errors, we have performed 
a bootstrap analysis, as described by \cite{efron82}. For our analysis we selected 
$25\,000$ random samples with replacement (`bootstrap samples') of our flux measurements 
and fit each sample to form a probability distribution of the resulting parameters. 
From this we measured 
a best-fitting gradient and $68$ per cent-confidence range of 
$0.015 \pm {0.003 \atop 0.002}$~Jy~yr$^{-1}$.
This analysis allows us to rule out a flat or decreasing flux density (a line with gradient $\le0$) 
with $99.95$ per cent confidence.
\changes{We also constructed light curves for several field sources common to most of the
epochs in our sample. 
There was no systematic change in flux density over this period of our observations.}
Comparison with our fitted flux value of 1.23 Jy on 2005/01/01 gives a percentage increase of
$1.22$ per cent yr$^{-1}$ with $68$ per cent-confidence range of $1.06 - 1.46$ per cent yr$^{-1}$.

\begin{figure}
\begin{center}
\includegraphics[width=\linewidth]{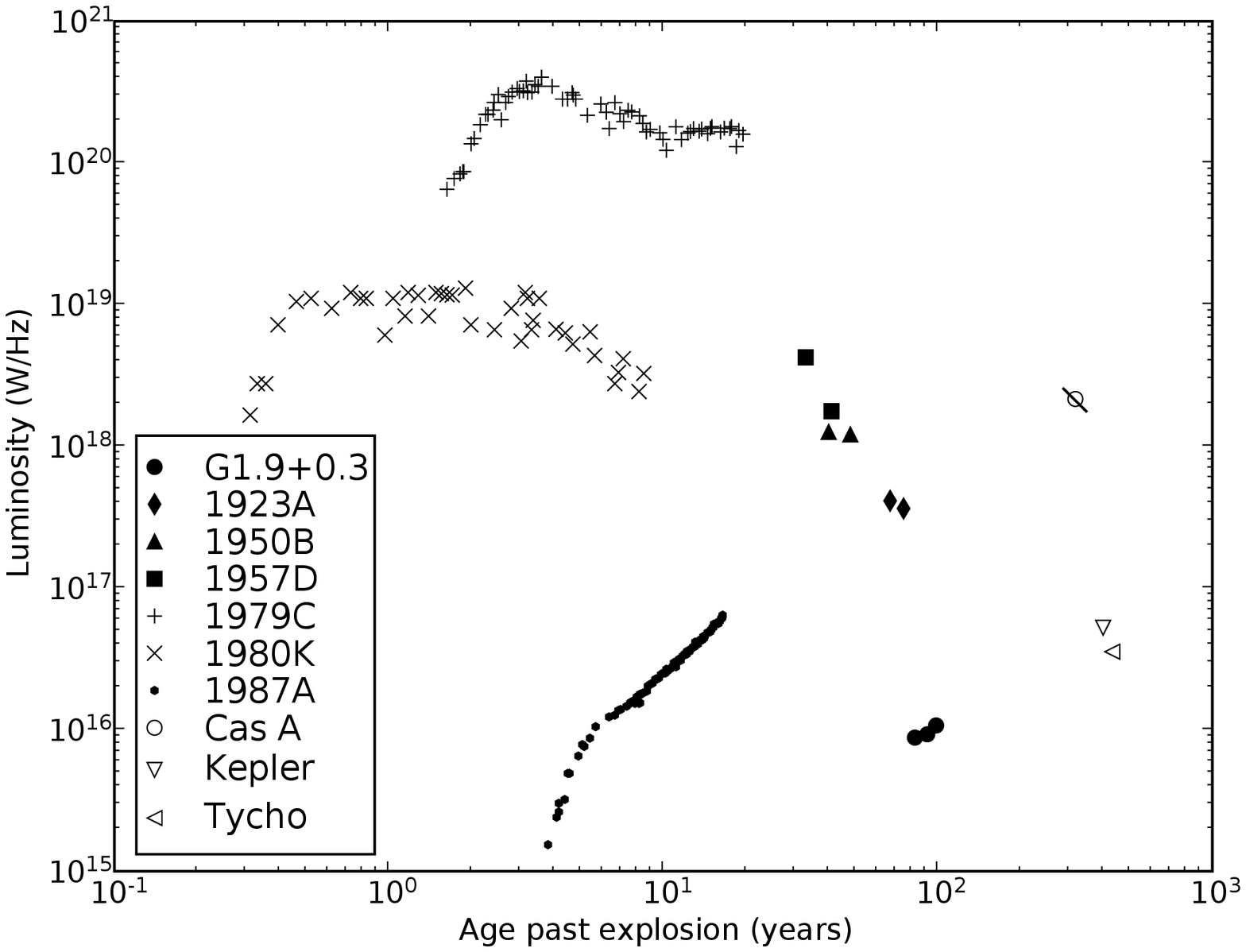}
\caption{Comparison of the $\sim1.4$ GHz light curve of \sng\ with those of other RSNe and 
SNRs of known ages. The MOST flux densities of \sng\ have been scaled to 1.4 GHz assuming 
a spectral index of $\alpha = 0.63$. For clarity, the MOST data have been binned into three 
bins corresponding to the three groups of measurements ($\sim1990$, $\sim1998$, $\sim2005$).
The RSNe are SN 1923A, SN 1950B, SN 1957D \citep[all from][]{stockdale06}, SN 1979C \citep{weiler91,montes00}, SN 1980K \citep{weiler86,montes98} and SN 1987A \citep{staveley-smith07}. The SNRs are Cas A \citep[with the $1965-1999$ fading shown, extended for clarity;][]{reichart00}, Kepler (SN 1604) and Tycho (SN 1572) \citep[both from][]{green04}.
}\label{f_lumcmp}
\end{center}
\end{figure}

Taking the age estimate of $\sim100$ years from 
\citet{green08}, we have plotted the light curve of \sng\ in Fig.~
\ref{f_lumcmp} with a selection of other RSNe (SN 1923A, SN~1950B, 
SN 1957D, 1979C, 1980K) and SNRs (Cas A, Kepler -- SN 1504, Tycho -- SN 1572)
for comparison.
To convert flux density to luminosity, we assumed a distance 
of 8.5 kpc for \sng, following the argument of \citet{reynolds08} that the high 
absorption towards \sng\ makes it unlikely that it is significantly closer 
than the Galactic Center. For plotting clarity, the points have been binned into
three groups of measurements ($\sim1990$, $\sim1998$, $\sim2005$).

\section{Discussion}
\label{s_discuss}
Only a fraction of SNe have detectable radio emission. Observations of these RSNe 
\citep[see, for example;][]{eck02} show that radio emission caused by the interaction 
between the shock and the circumstellar medium of the progenitor star is first detected 
within days to months of the initial explosion.
As the shock travels into regions of decreasing opacity, the radio emission brightens,
peaking between days to years after the explosion. 
After reaching their maximum brightness, RSNe then show a 
power law decrease in flux density with time \citep{weiler02}.
This is demonstrated by the light curves of SN~1923A, SN~1950B, and others, 
shown in Fig.~\ref{f_lumcmp}. The radio behaviour of RSNe is explained well by the 
models of \citet{chevalier82,chevalier98}.

There are even fewer Galactic SNRs of known age. Cassiopeia A,
the youngest SNR of known age at $\sim330$ years, 
Kepler (SN~1604) and Tycho (SN~1572) are shown in Fig.~\ref{f_lumcmp}.
The radio emission we detect from SNRs is due to the interaction between the 
shock and the interstellar medium.
Hence the radio emission from a young SNR will increase once it has swept up enough
of the surrounding interstellar medium. The timescale of this brightening is predicted 
to be $\sim100$ years \citep{gull73,cowsik84}. 

Between the youngest SNR of known age (Cas A) and the oldest RSNe of known age 
(SN~1923A), there is a gap in our observational evidence which makes it hard to probe
the period after the fading of the radio emission from the RSN and before the 
SNR switches on. \citet{eck02} conducted a search for radio emission from SNe of a range
of known ages, but did not detect any of the SNe that occurred prior to SN~1923
(for example, SN~1885A and SN~1909A). If the estimated age of $100-120$ years for \sng\ is 
correct, then our measurements help constrain the flux density evolution in this 
intermediate time range.

To explain the observed expansion rate measured by \cite{reynolds08} and the new light
curve presented here, we need to reconsider some of the standard assumptions made when
predicting the luminosity evolution of SNRs.
The radio luminosity $L_\nu$ of a synchrotron source at a given frequency $\nu$ is a function of 
the energy spectrum of the ultrarelativistic electrons, the magnetic field $B$
present in the source, and the volume of the source $V$, as given 
in \cite{longair94}
\[
L_\nu = A(\alpha)V\kappa B^{1+\alpha}\nu^{-\alpha} \; ,
\]
where $\alpha$ is the spectral index, $A(\alpha)$ is a constant, and $\kappa$ is 
defined in terms of the electron energy spectrum per unit volume
\[
N(E)  dE = \kappa E^{-(2\alpha + 1)} dE \; .
\]
Hence for a given frequency
\[
  L_\nu \propto V \kappa B^{1+\alpha} \; .
\]
If we make the assumptions that the expansion of the remnant is adiabatic, and that
magnetic flux freezing is applicable, then 
the magnetic field strength decreases as $B\propto r^{-2}$. 
Following \cite{longair94} we can derive that $V\kappa \propto r^{-2\alpha}$ and so
\[
  L_\nu \propto r^{-2(2\alpha+1)} \; .
\]
Now the time dependence of the luminosity is proportional to the time dependence 
of the source radius, since the spectral index $\alpha$ should not change during 
adiabatic expansion.
Expressing the change in radius with time as a power law with expansion parameter
$m$, $r\propto t^m$, then
\[
L_\nu \propto t^{-2m(2\alpha + 1)} \; .
\]
During an ideal free expansion phase $r\propto t$, and in the Sedov phase 
$r\propto t^{0.4}$. Using the spectral index of $\alpha = 0.63$ given by 
\cite{reynolds08} we would expect the luminosity to change as 
$L_\nu\propto t^{-4.5}$ in free expansion, or $L_\nu\propto t^{-1.8}$ in Sedov expansion. 

If we fit the measured expansion rate given by \cite{reynolds08} with a power law
(assuming an age of 100 years in 2008), we get $r\propto t^{0.55}$, implying $L_\nu\propto t^{-2.5}$ which is intermediate between free expansion and the Sedov phase.
Using our new light curve data we get $L_\nu\propto t$.
Clearly, to produce the observed increase in luminosity, either the
magnetic field strength or the energy density of the relativistic electrons (or both)
must be increasing with time, rather than decreasing as assumed above. 
\changes{
A similar conclusion was drawn by \citet{green08}
}. 
This scenario is supported by what we know about older SNRs ---
extrapolating backwards from the observed magnetic field strength in Cas A, the assumption
of magnetic flux freezing would result in an implausible magnetic field strength in the 
progenitor star \citep{longair94}. Hence at some period in its evolution, the magnetic 
field strength must have undergone a process of amplification. Simulations by \citet{jun99} 
show that a possible mechanism for this amplification is Rayleigh-Taylor instability in
the interaction region between the SNR shock and a surrounding cloud of interstellar gas.
This happens on timescales of $\sim100-300$ years after the inital explosion.
Our observations suggest that \sng\ has entered such a phase, in which $\kappa B$ is 
growing.

\section{Summary}
Twenty years of observations with the Molonglo Observatory Synthesis Telescope
show that the young supernova remnant \sng\ has increased in brightness
by $1.22 \pm {0.24 \atop 0.16}$ per cent yr$^{-1}$ between 1998 and 2007. This supports the estimate of 
$\sim2$ per cent yr$^{-1}$ made by \cite{green08}. This result, for what is potentially
the youngest SNR of known age at $\sim100$ years, confirms predictions that 
although most RSNe are decreasing in brightness, they will have to go through a 
period of brightening later in their evolution, to match the observed brightness of
young SNRs in our Galaxy.

As the next generation of radio telescopes revolutionise the study of RSNe and SNRs,
archival data from telescopes such as MOST will have an important 
role to play, giving us insight into the evolution of these objects over long timescales.

\section*{Acknowledgments} 
Thanks to D. A. Green and colleagues for making their VLA images available.
We thank B. Piestrzynska for processing the MOST archival
data and to the MOST site manager, 
D. Campbell-Wilson.
The MOST is operated with the support of the Australian Research 
Council and the School of Physics, The University of Sydney.
TM and BMG acknowledge the support of an ARC Australian Postdoctoral 
Fellowship (DP0665973) and an ARC Federation Fellowship (FF0561298), respectively.
SC acknowledges support from the University of Sydney Postdoctoral Fellowship program.

\label{lastpage}
\end{document}